\documentclass[12pt,preprint]{aastex}
\usepackage{psfig}
\usepackage{color}

\def\hnot{\ifmmode H_0 \else H$_0$ \fi}

\def\msun{\ifmmode {\rm M_\odot} \else M$_\odot$\fi}
\def\lsun{\ifmmode {\rm L_\odot} \else L$_\odot$\fi}

\def\deg{\ifmmode ^{\circ}
         \else $^{\circ}$\fi}
\def\pdeg{\ifmmode 
           $\setbox0=\hbox{$^{\circ}$}\rlap{\hskip.11\wd0 .}$^{\circ}
     \else \setbox0=\hbox{$^{\circ}$}\rlap{\hskip.11\wd0 .}$^{\circ}$\fi}

\def\arcsec{\ifmmode '' \else $''$\fi}
\def\arcsecpt{\ifmmode ''\!\!. \else $''\!\!.$\fi}

\def\msunyr{\ifmmode {\rm M_\odot~yr^{-1}}\else${\rm M_\odot~yr^{-1}}$\fi}
\def\lam{\ifmmode {\lambda} \else {$\lambda$} \fi}

\def\mdoto{\ifmmode {\dot{M}_0} \else  $\dot{M}_0$ \fi}
\def\teff{\ifmmode {T_{eff}} \else $T_{eff}$ \fi}
\def\ilam{\ifmmode {I_\lambda} \else  $I_\lambda$ \fi}
\def\inu{\ifmmode {I_\nu} \else  $I_\nu$ \fi}
\def\fnu{\ifmmode {F_\nu} \else  $F_\nu$ \fi}
\def\lhost{\ifmmode {L_{\rm host}} \else  $L_{\rm host}$\fi}

\def\yr{\ifmmode {\rm yr} \else  yr \fi}
\def\cm{\ifmmode {\rm cm} \else  cm \fi}
\def\cmmitwo{\ifmmode \rm cm^{-2} \else $\rm cm^{-2}$\fi}
\def\cmmithree{\ifmmode \rm cm^{-3} \else $\rm cm^{-3}$\fi}
\def\cmps{\ifmmode \rm cm~s^{-1}\else $\rm cm~s^{-1}$\fi}
\def\cmpsps{\ifmmode \rm cm~s^{-2}\else $\rm cm~s^{-2}$\fi}
\def\kmps{\ifmmode \rm km~s^{-1}\else $\rm km~s^{-1}$\fi}
\def\kmpspmpc{\ifmmode \rm km~s^{-1}~Mpc^{-1} \else
    $\rm km~s^{-1}~Mpc^{-1}$\fi}
\def\ergps{\ifmmode \rm erg~s^{-1} \else $\rm erg~s^{-1}$ \fi}
\def\ergpspcm{\ifmmode \rm erg~s^{-1}~cm^{-2} \else $\rm erg~s^{-1}~cm^{-2}$ \fi}
\def\ergpspcmphz{\ifmmode \rm erg~s^{-1}~cm^{-2}~Hz^{-1} \else $\rm
erg~s^{-1}~cm^{-2}~Hz^{-1}$ \fi}
\def\ergpspcmpa{\ifmmode \rm erg~s^{-1}~cm^{-2}~\AA^{-1} \else $\rm
erg~s^{-1}~cm^{-2}~\AA^{-1}$ \fi}
\def\ergpsphz{\ifmmode \rm erg s^{-1} Hz^{-1} \else 
   $\rm erg s^{-1} Hz^{-1}$ \fi} 

\def\mhostsigthree{$M_{\rm{HOST}} - \sigma_{[O~III]}$}
\def\mbh{\ifmmode M_{\rm{BH}} \else $M_{\mathrm{BH}}$ \fi}
\def\mhost{\ifmmode M_{\rm{{HOST}}} \else $M_{\rm{{HOST}}}$ \fi}
\def\mbhsig{$M_{\rm{BH}} - \sigma$}
\def\mbhbulge{\ifmmode M_{\rm{BH}} - M_{\rm{bulge}} \else $M_{\rm{BH}}
- M_{\rm{bulge}}$\fi}

\def\mbhsigstar{\ifmmode M_{\rm{BH}} - \sigma_* \else $M_{\rm{BH}} -
\sigma_*$ \fi}
\def\mbhsigthree{\ifmmode M_{\rm{BH}} - \sigma_{\rm{[O~III]}} \else $M_{\rm{BH}} - \sigma_{\rm{[O~III]}}$ \fi}

\def\sigstar{\ifmmode \sigma_* \else $\sigma_*$ \fi}
\def\sigthree{\ifmmode \sigma_{\rm{[O~III]}} \else $\sigma_{\rm{[O~III]}}$ \fi}
\def\wthree{\ifmmode {\rm FWHM([O~III])} \else $\rm{FWHM([O~III])}$ \fi}
\def\mthree{\ifmmode M_{\rm{[O~III]}} \else $M_{\rm{[O~III]}}$ \fi}
\def\hbeta{\ifmmode {\rm H}\beta \else H$\beta$ \fi}
\def\oiii{[\ion{O}{3}]}
\def\feii{\ion{Fe}{2}}

\def\sref#1{\S~\ref{#1}}
\def\fref#1{Figure~\ref{#1}}

\def\be{\begin{equation}}
\def\ee{\end{equation}}

\def\eg{e.g.}

\accepted{}
\journalid{}{}
\articleid{}{}

\slugcomment{Accepted by ApJ}

\shortauthors{E. Bonning}
\shorttitle{Magnitude - [OIII] width}
\received{some future date}
\begin{document}

\title{QSO Narrow [OIII] Line Width and Host Galaxy Luminosity}

\author{
E. W. Bonning \altaffilmark{1},
G. A. Shields\altaffilmark{2},
S. Salviander\altaffilmark{2},
R. J. McLure \altaffilmark{3}
}

\altaffiltext{1}{ Laboratoire de l'Univers et de ses Th\'{e}ories,
  Observatoire de Paris, F-92195 Meudon Cedex, France; erin.bonnning@obspm.fr} 

\altaffiltext{2}{Department of Astronomy, University of Texas at
 Austin, Austin, TX 78712}

\altaffiltext{3}{Institute for Astronomy, University of Edinburgh,
  Royal Observatory, Edinburgh EH9 3HJ}

\begin{abstract}

Galaxy bulge luminosity 
$L$, black hole mass \mbh, and stellar velocity 
dispersion \sigstar  increase together in a way suggesting a close
evolutionary relationship. Measurements of the \mbhsigstar
relationship as a function of cosmic time may shed light on 
the origin of this relationship.  Direct measurements of $\sigstar$ at
high redshift are difficult, and the width of the narrow emission lines
of AGN has been proposed as a surrogate for $\sigstar$.  We investigate
the utility of  using  \sigthree\  for \sigstar\ by 
examining host galaxy magnitudes and \oiii\ line widths for low 
redshift QSOs. For  radio-quiet QSOs, $\sigthree$
is consistent in the mean with the value of $\sigstar$ predicted by
the Faber-Jackson relation.  For our limited range of \lhost, scatter
obscures the expected increase of \sigthree\ with \lhost.   However,
for a sample of AGN covering a wide range of measured or inferred
\sigstar, there is a clear increase of \sigthree\ with
\sigstar. Radio-loud QSOs on average  have \sigthree smaller by
$~0.1$ dex than radio-quiet QSOs of similar \lhost, at least for
luminosities typical of PG QSOs.  Star formation rates in our low
redshift QSOs are  smaller than required to maintain the typical
observed ratio of bulge mass to black hole mass.

\end{abstract}


\section{INTRODUCTION}
\label{sec:intro}

The relationship between a galaxy's central  black hole and the
evolutionary history of its host galaxy is unclear.
Probing the correlations between galaxy luminosity,
stellar velocity, gas velocity, and black hole mass may reveal the
connections between a galaxy's central black hole and its formation
history. 

Galaxy magnitudes are closely related to the mass of the central black
hole (\mbh). {Kormendy} \& {Richstone} (1995) and {Magorrian} {et~al.} (1998) 
showed that the central black hole mass correlates with  the bulge
mass and luminosity. 
{Laor} (1998) showed that quasar host galaxy luminosity is similarly
correlated with the black hole mass deduced from the broad $\hbeta$ lines.
The stellar velocity dispersion (\sigstar) in the galactic bulge also
correlates with 
the mass of the central black hole. {Gebhardt} {et~al.} (2000a) and
{Ferrarese} \& {Merritt} (2000) found that the correlation between \mbh and
\sigstar is strong, suggesting a link between the formation of
the bulge and the black hole. Theoretical interpretations of
this correlation (\eg~ {Silk} \& {Rees} 1998; {Adams}, {Graff}, \& {Richstone} 2001; {Burkert} \& {Silk} 2001; {Ostriker} 2000; {Balberg} \& {Shapiro} 2002; {Haehnelt} \& {Kauffmann} 2000; {Murray}, {Quataert}, \&  {Thompson} 2004)
differ as to whether the black hole forms before, during, or after the
bulge.  Measurements of the \mbhsigstar\ relationship at high
redshift may help to resolve this question ({Shields} {et~al.} 2003).

Derivation of \mbh\ from AGN  broad line widths and continuum
luminosity is now well established
(see {Kaspi} {et~al.} 2000; {McLure} \& {Dunlop} 2004, and references therein). 
In contrast, measurements of \sigstar\ are difficult for distant QSOs, given
the faintness of the galaxy and the relative brightness of the nucleus.
However, it may be possible to infer $\sigstar$ from the
widths of the narrow emission lines.
{Nelson} \& {Whittle} (1996) made a comparison of bulge magnitudes, \oiii\
$\lambda\lambda 5007, 4959$ line widths and stellar velocity dispersions
in Seyfert galaxies, taking $\rm{\sigthree \equiv FWHM{[O~III]}}/2.35$
as appropriate for a Gaussian line profile.
They found on average good agreement between \sigthree and \sigstar,
although \sigthree\ shows more scatter than \sigstar on a Faber-Jackson
plot.  This supports the idea that the NLR gas is largely in orbital
motion in the gravitational potential of the bulge and can be effectively
used as a substitute where stellar velocities cannot be
measured. Further supporting the use of \sigthree for \sigstar is the work
of {Nelson} (2000), who shows that the \mbhsigstar
relation for normal galaxies and AGN 
({Gebhardt} {et~al.} 2000a, 2000b) is preserved when $\sigthree$ is used
in place of $\sigstar$.  

{Shields} {et~al.} (2003) examined the
\mbhsigthree\ relationship in QSOs with redshifts up to $z = 3$, finding
little evolution with cosmic time.  This result suggests that
supermassive black 
holes and their host galaxies grow together, or that both have largely
completed their growth by $z \approx 2$.  Such a conclusion would provide
valuable guidance to theories of the evolution of galaxies and their
black holes.  However, the use of \sigthree for \sigstar is controversial.
\oiii\ line profiles often have
substantial asymmetry and a non-Gaussian profile. This may
arise from outflow combined with extinction of the far side of the
NLR  (e.g., Wilson \& Heckman 1985; Nelson \& Whittle 1995). 
Objects with strong iron emission can obscure the \oiii\
emission due to the \feii\ features lying close to the
\oiii$\lambda\lambda 4959,5007$ line. 
These complications and no doubt others contribute to the large
scatter shown by \oiii\ emission in comparison to $\sigstar$. This
underscores the importance of quantifying the correspondance of
\sigstar with \sigthree  at QSO luminosities.

Direct comparisons of \sigthree with \sigstar have generally been limited to
lower luminosity AGN ({Smith}, {Heckman}, \& {Illingworth} 1990; {Nelson} \& {Whittle} 1996; {Onken} {et~al.} 2004). 
Indirect comparisons using the \mbhsigstar\ relationship
({Shields} {et~al.} 2003; {Boroson} 2003) rely on the
derivation of \mbh from the broad line widths.  Therefore it is
important to evaluate the substitution of \sigstar for \sigthree  as
directly as possible for QSO luminosities that more  closely approach the
luminosities of observed high redshift QSOs.  
In normal galaxies, \sigstar is related to bulge luminosity by the
Faber-Jackson relation 
({Forbes} \& {Ponman} 1999; {Kormendy} \& {Illingworth} 1983).  This relation is particularly true for
early-type galaxies, which comprise the majority of the hosts of
luminous quasars with $M_R < -24$ ({Dunlop} {et~al.} 2003; {Schade}, {Boyle}, \& {Letawsky} 2000). Additionally, {Woo} {et~al.} (2004)  have shown that
 the host galaxies of BL Lac objects
and radio galaxies  up to redshift $z \sim 0.34$ lie on the normal
 galaxy fundamental plane.  This allows an indirect determination of
 \sigstar for comparison with 
\sigthree, as done for Seyfert galaxies by {Nelson} \& {Whittle} (1996). 
In this paper, we test the use of
\oiii\ line widths as a  surrogate for \sigstar\ by studying the
\mhostsigthree\  relationship in a sample of quasars for which the
host galaxy luminosity has been measured.
In \sref{sec:data} we describe the host galaxy luminosities used here
and our measurements of the \oiii\ line widths. In \sref{sec:results}
we show that \sigthree agree closely with \sigstar in the mean, examine
the scatter in this agreement, and show that
\sigthree does indeed track
\sigstar\ over a broad range of QSO luminosities.  We assume a
cosmology with
$\hnot = 70~\kmpspmpc, \Omega_{\rm M} = 0.3, \Omega_{\Lambda} = 0.7$.
All values of luminosity used are corrected to these cosmological parameters.

\section{DATA}
\label{sec:data}

\subsection{Host Galaxy Magnitudes}
\label{subsec:mag}

Host galaxy magnitudes for ellipticals, and bulge magnitudes
for spiral hosts were taken from the literature, including 
 {McLure} \& {Dunlop} (2002),  {Percival} {et~al.} (2001),
{Floyd} {et~al.} (2004), {McLure}, {Percival}, \& {Dunlop} (2004) and
{Hamilton}, {Casertano}, \&  {Turnshek} (2002). The host galaxy measurements in the first four papers
(i.e., excepting {Hamilton} {et~al.}) were performed using the same 
method for fitting and subtracting the nucleus, modeling the
host galaxy, integrating  host galaxy light, and performing 
K-corrections, evolution corrections, and correction for Galactic
extinction  (described in {McLure}, {Dunlop}, \& {Kukula} 2000).  
The measurements done by {Hamilton} {et~al.} (2002) differed in these
respects. As a result, when comparing the set of objects for which
both sudies report a host galaxy magnitude,
there is an average offset.  We converted the V-band
magnitudes of {Hamilton} {et~al.} to the Cousins R-band in which 
{McLure} \& {Dunlop} report their results by using colors $V - R_c =
0.61$ for elliptical  
galaxies and  $V - R_c = 0.54$ for spiral galaxies ({Fukugita}, {Shimasaku}, \&  {Ichikawa} 1995),
taking the morphologies as given in {Hamilton} {et~al.} The resulting 
magnitudes are, on average, brighter than those of {McLure} \& {Dunlop} (2002) by
about 0.25 mag. We therefore 
adjusted all magnitudes measured by {Hamilton} {et~al.} by this
amount in order to correct for average measurement difference 
between the two methods.

\subsection{\oiii\ Line Widths}
\label{subsec:oiiiwidth}

The
\oiii\ line widths in this paper were measured 
directly from spectra publicly
available from {Marziani} {et~al.} (2003) and the spectra presented by
{McLure} \& {Dunlop} (2001). Line widths were corrected by subtracting in
quadrature the instrumental FWHM, which 
ranged from 3 - 7~\AA. Iron emission was subtracted from the spectra
using the
{Boroson} \& {Green} (1992) \feii\ template so as to eliminate the emission
bands around $\lambda4500$ and $\lambda5200$.  Objects for which
\feii\ emission obscured the \oiii\ line were discarded. For the
{McLure} \& {Dunlop} sample, this led to the exclusion
of one out of 13 radio-loud QSOs (RLQ) and 2 out of 17 radio-quiet
QSOs (RQQ).  Of the 7 objects from McLure,
Percival, \& Dunlop, all RLQ, one was rejected.  Of the three in the
{Floyd} {et~al.} sample that had spectra from Marziani et al., one
was rejected leaving 1 RQQ and 1 RLQ.  Of the objects from Hamilton et al.
with available spectra, four duplicated objects already in the foregoing
sources, and one was rejected, leaving 3 RLQ and 8 RQQ.

We made a direct measurement of
the FWHM using the IRAF
\footnote{IRAF is distributed by the National Optical Astronomy
Observatories,
    which are operated by the Association of Universities for Research
    in Astronomy, Inc., under cooperative agreement with the National
    Science Foundation.}
routine SPLOT, in preference to using a fit to the line
profile. Typical errors in  FWHM(\oiii) are about 10\%, coming largely
from uncertainy in continuum placement.  
Objects with W(\feii\ ) $>$ 50~\AA\ have FWHM \oiii\ about 0.1 dex
wider than in objects with weaker \feii\ (Salviander, Shields, \& Gebhardt
2005, in preparation). This has the potential to  bias our sample
towards objects with narrow \oiii\, since we must discard objects in
which the \oiii\ is obscured by strong \feii\ emission.  However,
we only discard about one in ten objects, giving a negligible 0.01
dex effect on the mean values. FWHM \oiii\ are converted to velocity
dispersion as \sigthree\ by $\rm{\sigthree \equiv FWHM{[O~III]}}/2.35$
following {Nelson} (2000) and others.

\subsection{Black Hole Masses}
\label{subsec:blackholes}

Below we compare host galaxy magnitudes and \oiii\ widths with
black hole masses for our objects.  Black hole masses were derived
from the FWHM of the broad component of the \hbeta\ emission line and
the continuum luminosity at 5100 \AA, using equation A7 of
{McLure} \& {Dunlop} (2004, see references therein for details and background). 
Objects from {McLure} \& {Dunlop} (2002) had \hbeta\ width and continuum
luminosity given in {McLure} \& {Dunlop} (2001).
For the other objects, \hbeta\ width and the continuum
luminosity at $\lambda5007$ were taken from {Marziani} {et~al.},
with the  $\lambda5007$ scaled to a $\lambda5100$ value
using a typical continuum slope
$F_\lambda \propto \lambda^{-1.5}$. Two objects (0204+292 and
2247+140) were rejected because of weak or asymmetrical \hbeta, a
strong narrow component of \hbeta, and noise; and 0100+020 was omitted
because {Marziani} {et~al.} do not give an absolute flux.  The black
hole masses here are  typically several tenths dex smaller than those
tabulated by {McLure} \& {Dunlop} (2001) because of differences in the adopted
cosmology and mass formula.

\begin{table*}
\caption{Below are the objects for which we were able to obtain both host
  magnitudes and reliable \oiii\ line widths. Host galaxies are given
  in the Cousins R-Band and adjusted for the adopted cosmology, and
 $ \sigthree = \rm{FWHM{[O~III]}}/2.35$. Magnitude sources: 
(1) {McLure} \& {Dunlop} (2002, see references therein); 
(2) {Hamilton} {et~al.} (2002); 
(3) {Floyd} {et~al.} (2004);
(4)  Percival et al. (2001)
(5) {McLure} {et~al.} (2004)
\oiii\ sources: (a) {McLure} \& {Dunlop} (2001); (b) {Marziani} {et~al.} (2003). }
\begin{tabular}{lccccc}
\hline
\hline
Name & Redshift & $M_{R_c}$(host) & log($\sigma_{[O~III]}$) (km/s) &
Magnitude Source & \oiii\ source  \\
\hline
\multicolumn{6}{c}{Radio-Quiet Quasars} \\
0052+251 & 0.154 & -22.44 & 2.44 & (1) & (a) \\
0054+144 & 0.171 & -23.11 & 2.59 & (1) & (a) \\
0100+020 & 0.393 & -22.11 & 2.14 & (2) & (b) \\
0137-010 & 0.335 & -21.66 & 2.25 & (4) & (b) \\
0157+001 & 0.164 & -23.76 & 2.45 & (1) & (a) \\
0204+292 & 0.109 & -22.81 & 2.36 & (1) & (a)\\
0205+024 & 0.155 & -20.83 & 2.54 & (1) & (b)\\
0244+194 & 0.176 & -22.45 & 2.31 & (1) & (a)\\
0923+201 & 0.190 & -22.76 & 2.57 & (1) & (a) \\
0953+414 & 0.239 & -22.32 & 2.41 & (1) & (a) \\
1012+008 & 0.185 & -23.26 & 2.67 & (1) & (b) \\
1029-140 & 0.086 & -22.15 & 2.40 & (1) & (b) \\
1116+215 & 0.177 & -23.17 & 2.73 & (1) & (b) \\
1202+281 & 0.165 & -22.21 & 2.28 & (1) & (b) \\
1216+069 & 0.331 & -22.21 & 2.16 & (2) & (b) \\
1219+755 & 0.071 & -22.11 & 2.14 & (2) & (b) \\
1307+085 & 0.155 & -21.97 & 2.34 & (1) & (b) \\
1309+355 & 0.184 & -22.94 & 2.48 & (1) & (b) \\
1416-129 & 0.129 & -21.32 & 2.42 & (2) & (b) \\
1635+119 & 0.146 & -22.55 & 1.95 & (1) & (b) \\
1821+643 & 0.297 & -24.47 & 2.43 & (3) & (b) \\
\hline
\end{tabular}

\end{table*}
\begin{table*}
\begin{tabular}{lccccc}
\hline
\hline 
Name & Redshift & $M_{R_c}$(host) & log($\sigma_{[O~III]}$) &
Magnitude Source & \oiii\ Source  \\
\hline
\multicolumn{6}{c}{Radio-Loud Quasars}\\ 
0133+207 & 0.425 & -23.04 & 2.28 & (2) & (b) \\
0137+012 & 0.258 & -23.55 & 2.29 & (1) & (a) \\
0202-765 & 0.389 & -22.79 & 2.26 & (2) & (b) \\
0837-120 & 0.198 & -22.73 & 2.38 & (2) & (b) \\
1004+130 & 0.240 & -23.63 & 2.39 & (1) & (a) \\
1020-103 & 0.197 & -22.87 & 2.46 & (1) & (a) \\
1150+497 & 0.334 & -23.45 & 2.19 & (3) & (b) \\
1217+023 & 0.240 & -23.19 & 2.20 & (1) & (b) \\
1226+023 & 0.158 & -23.79 & 2.71 & (1) & (b) \\
1302-102 & 0.286 & -23.48 & 2.39 & (1) & (b) \\
1425+267 & 0.366 & -23.19 & 2.20 & (5) & (b) \\
1512+370 & 0.371 & -23.69 & 2.30 & (5) & (b) \\
1545+210 & 0.266 & -23.12 & 2.33 & (1) & (b) \\
1704+608 & 0.371 & -23.36 & 2.24 & (5) & (b) \\
2135-147 & 0.200 & -22.90 & 2.35 & (1) & (b) \\
2141+175 & 0.213 & -23.13 & 2.54 & (1) & (b) \\
2247+140 & 0.237 & -23.36 & 2.33 & (1) & (a) \\
2251+113 & 0.323 & -22.70 & 2.41 & (5) & (b) \\
2308+098 & 0.432 & -22.92 & 2.32 & (5) & (b) \\
2349-014 & 0.173 & -23.77 & 2.24 & (1) & (b) \\
2355-082 & 0.210 & -23.12 & 2.30 & (1) & (a) \\
\hline   
\end{tabular}
\end{table*}

\clearpage

\vspace{2.cm}
\begin{figure}[]
\begin{center}
\plotone{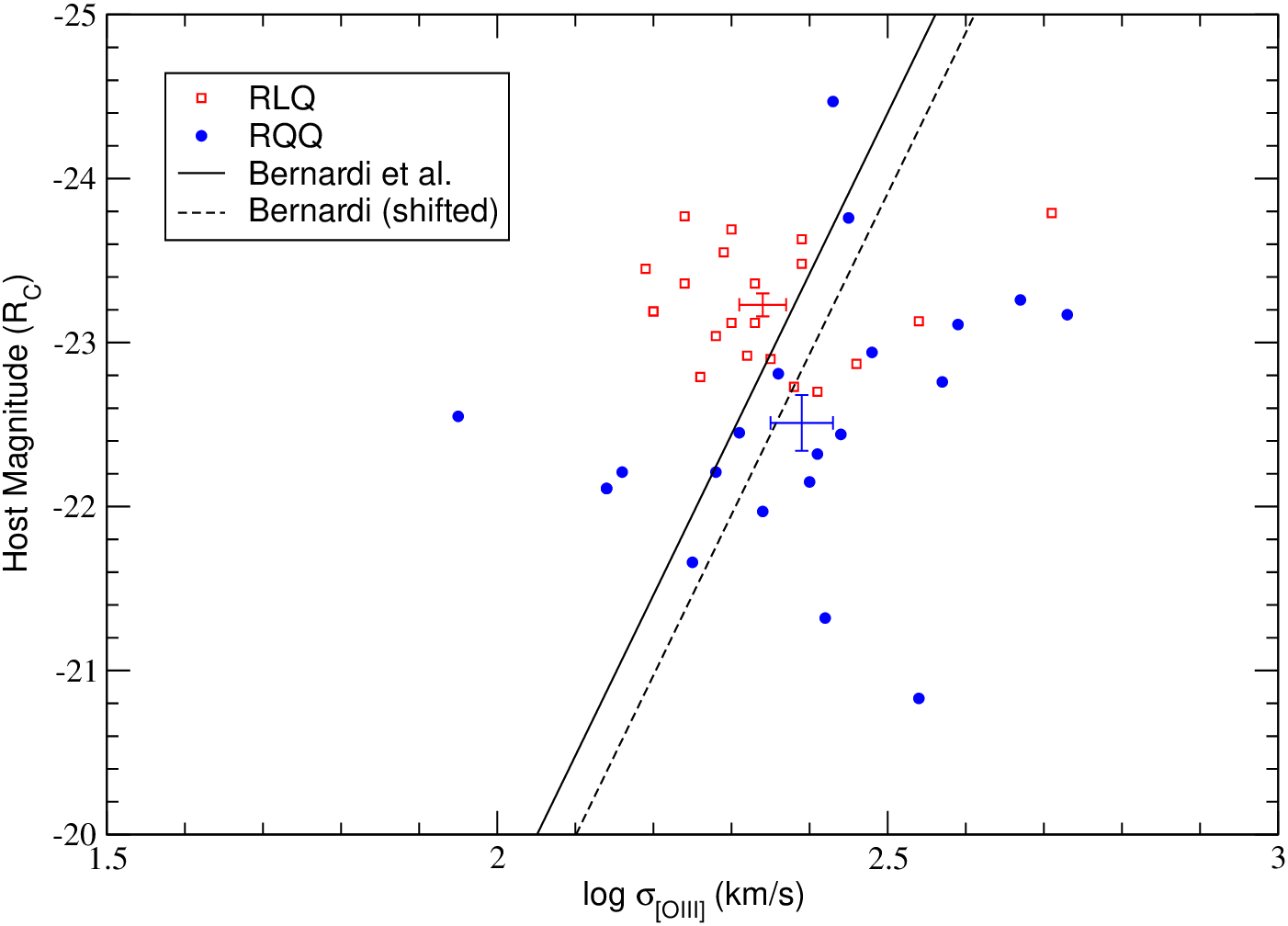}
\figcaption[fig1]{The above plot shows the sample of quasars for 
which we have host galaxy bulge
magnitudes and reliable \oiii\ FWHM. The objects were
classified as radio-loud or quiet according to the papers from which
bulge magnitudes were taken. The straight line is the Faber-Jackson
relation measured by {Bernardi} {et~al.} (2003a, 2003b); it is not a fit to the
data. The dashed line is the same relation with log $\sigma$
displaced by 0.05 to account for the remark by  {Bernardi} {et~al.} (2003b, p.1854) that their
meausured  \sigstar is smaller than the results of other authors by
this amount. The crosses indicate
the mean values and errors of the mean  for host luminosity and
$\sigthree$ for RL and RQ objects, the RL being above and to the left
of the RQ mean. \label{fig:plot1}}  
\end{center}
\end{figure}
\begin{figure}[]
\begin{center}
\plotone{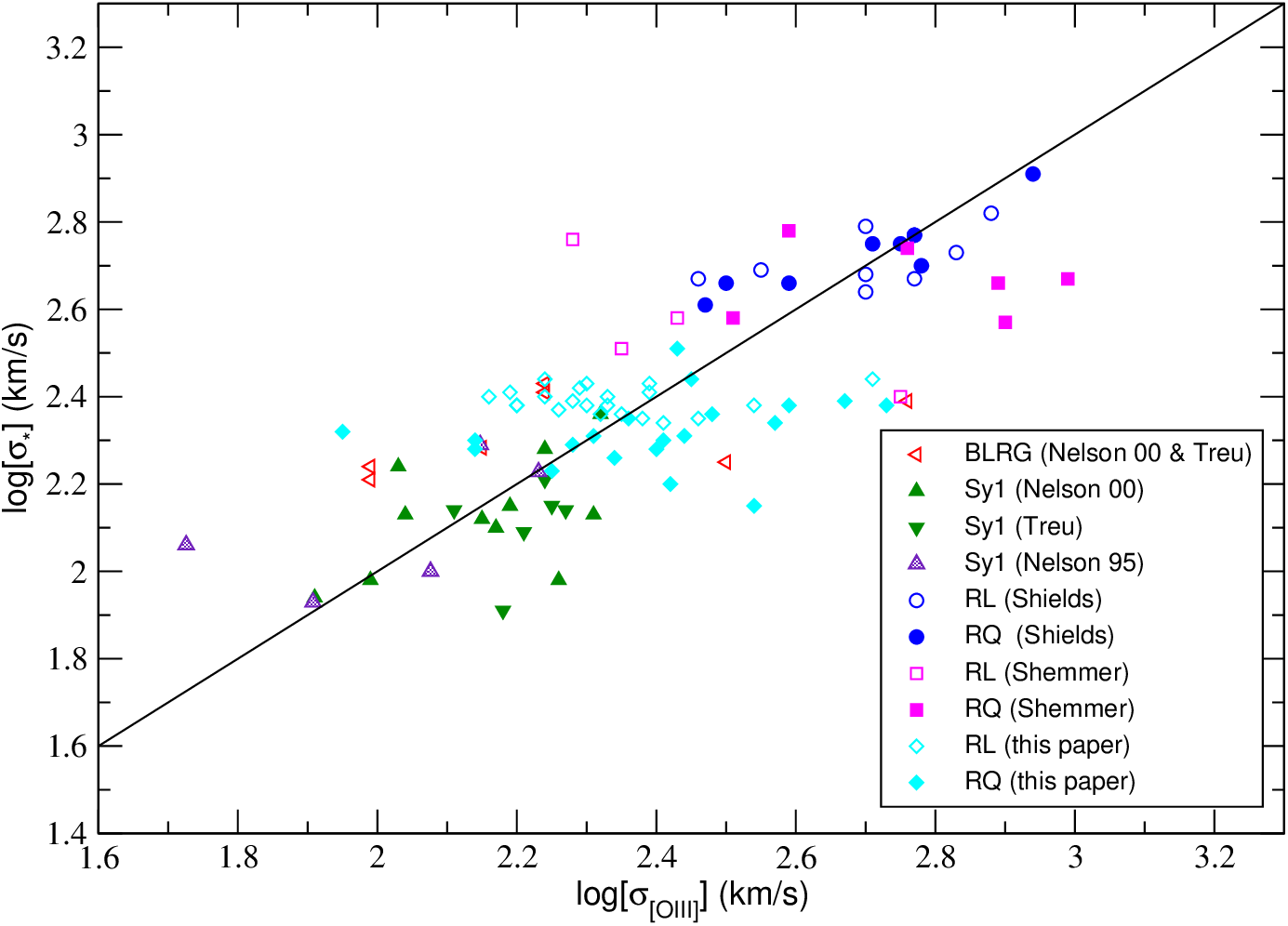}
\figcaption[fig2]{\sigthree\ v \sigstar\ (or a surrogate)  for our
  data sample and  others.\label{fig:sigsig} The broad line radio
  galaxies (BLRG) and  Seyferts are given with measured \sigstar. The
  QSOs from Shields et 
al. and Shemmer et al. have \sigstar calculated from \mbh. QSOs from
this paper have \sigstar calculated from their host galaxy
luminosity. Details of the data sets are given in the text. }
\end{center}
\end{figure}
\begin{figure}[]
\begin{center}
\plotone{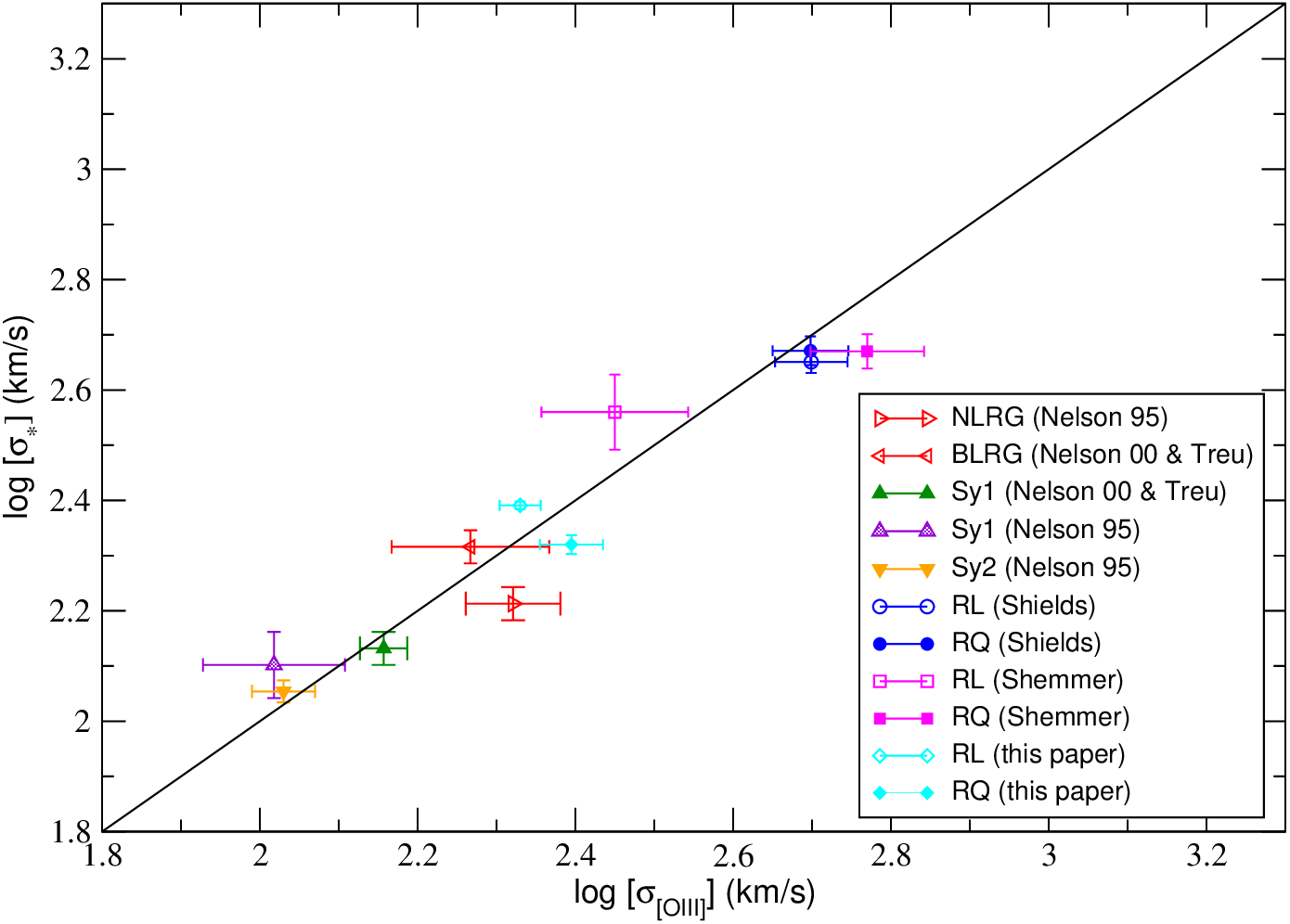}
\figcaption[fig3]{The means of each data set with error bars
  indicating the standard deviation of the mean. Also included are
  narrow line radio galaxies (NLRG) and Seyfert 2 objects from
  {Nelson} \& {Whittle} (1995).\label{fig:sigsigmeans}} 
\end{center}
\end{figure}
\begin{figure}[]
\begin{center}
\plotone{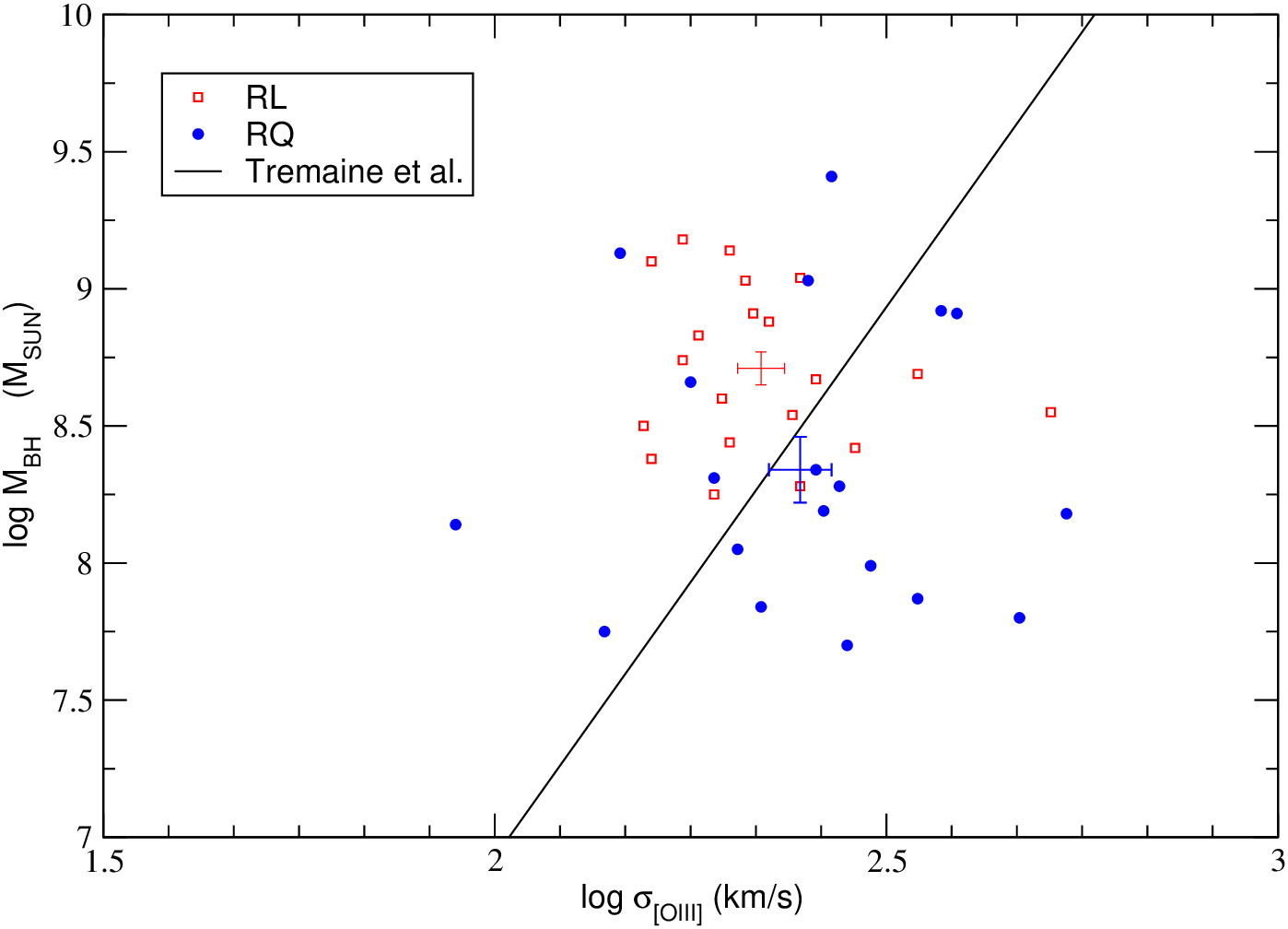}
\figcaption[fig4]{\mbh\ versus \sigthree for objects in Table 1 with
 exceptions described in the text, and expressed in log $M/\msun$. The
\sigthree\ values are taken from Table 1. The RL objects
are offset from the {Tremaine} {et~al.} (2002) relation similarly to the RL-RQ
offset in \fref{fig:plot1} \label{fig:plot2}}
\end{center}
\end{figure}

\begin{figure}[]
\begin{center}
\plotone{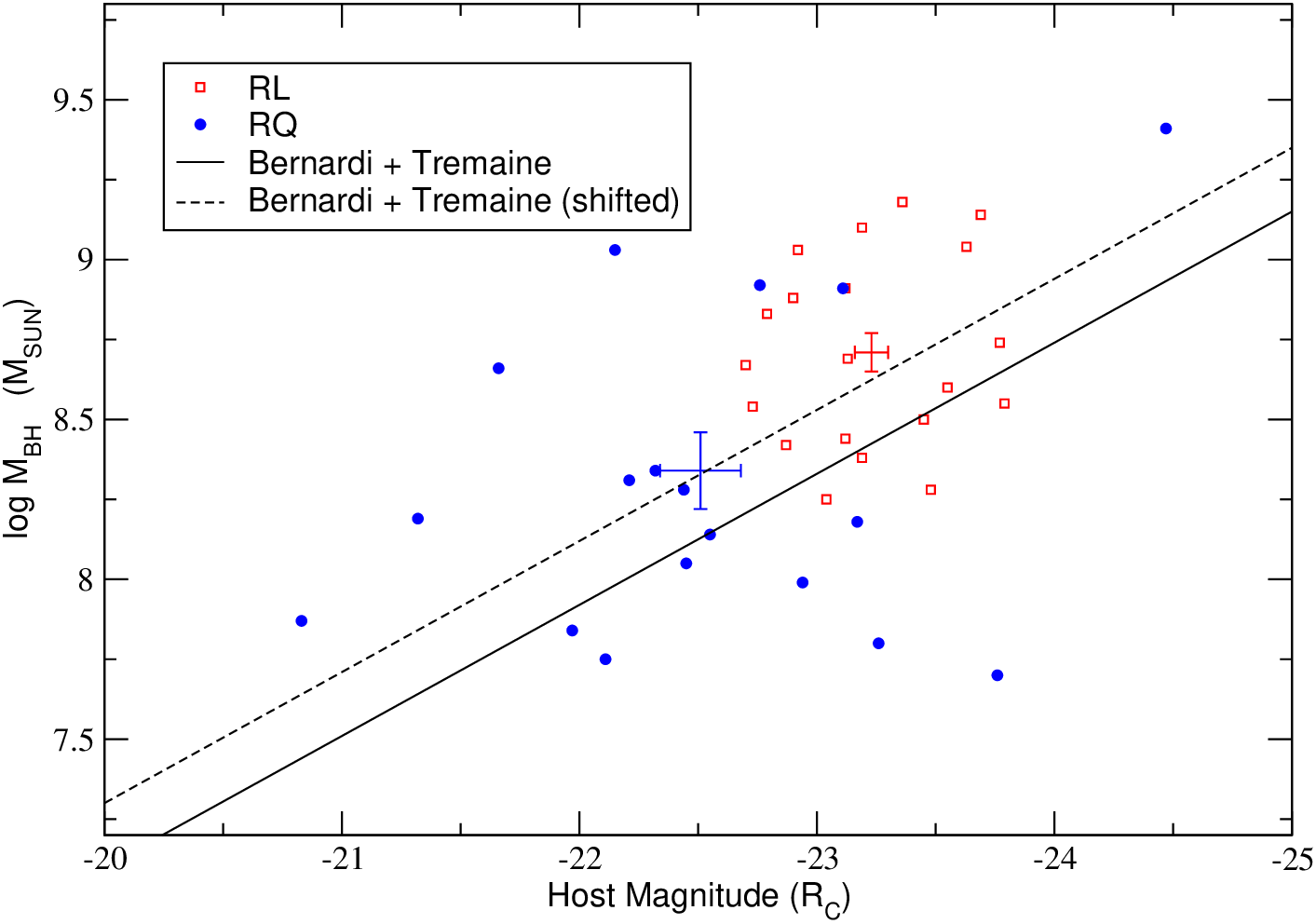}
\figcaption[fig5]{\mbh\ versus \mhost\  for the same objects as
\fref{fig:plot2}, along with the relation described in the
text. This figure is similar to  
Figure 2 of {McLure} \& {Dunlop} (2001), where it can be seen that the RL objects
are not offset from the RQ objects in relation to the normal
\mbh\ - \mhost\ trend. \label{fig:plot3}}
\end{center}

\end{figure}

\section{RESULTS}
\label{sec:results}

\subsection{The \mhostsigthree\ relationship}
\label{subsec:mhostsigthree}

Our results for \mhostsigthree\ are plotted in \fref{fig:plot1}. 
The Faber-Jackson relation is shown as taken from
{Bernardi} {et~al.} (2003a, 2003b) with $L \sim
\sigma^{3.91}$. {Bernardi} {et~al.} note that their measurements
of \sigstar are 0.05 dex smaller than those previously published. We also
indicate the relation with the changed zero point in \fref{fig:plot1}.
As can be seen, the data for the QSOs given agree in the mean with the 
Faber-Jackson relation, using $\sigthree$ in lieu of $\sigstar$. However, the
scatter is large, about 0.16 in log \sigthree, for the full sample
or 0.11 dex for the radio-quiet objects excluding the extreme outlier 
(1635+119) at low \sigthree. Given the limited range of luminosities
available to us this scatter obscures any increase of \sigthree\ with
\mhost as expected by the Faber-Jackson relation.

Scatter inherent in the Faber-Jackson relation for normal galaxies
contributes very little to the scatter seen in \fref{fig:plot1}. 
{Bernardi} {et~al.} (2003a, 2003b), in a sample of 9000 early-type
galaxies, find the rms 
scatter in $\sigstar$ in the R-band to be around .075 dex in log
$\sigma$ at constant luminosity (see Figure 4 in {Bernardi} {et~al.} 2003b).
Another  source of uncertainty comes from the inability to
measure FWHM \oiii\ with high accuracy. Plausible displacements in location
of the continuum level can lead to a $\Delta \wthree$ of about $\pm 50~
\rm km/s$, or a scatter of $\sim 0.04$ dex in log
\sigthree. Measurement error of $\sim$ 0.5 mag. in \mhost corresponds
to 0.05 in log \sigthree.
Taken together with  the standard deviation in log \sigthree\ of $\sim
0.16$, one is left with an intrinsic scatter in \oiii\ of

\begin{equation}
({\rm log}~\sigthree)^2 \sim (0.16)^2_{OBS} - (0.075)^2_{FJ} -
  (0.04)^2_{FWHM} - (0.05)^2_{M_{HOST}} = (0.13)^2
\end{equation}

Clearly, the geometry and kinematics of the NLR cause the \sigthree\ to
differ substantially from \sigstar\ in individual objects. However,
some mix of processes evidently increases or decreases \sigthree\ with
respect to \sigstar\ with roughly equal probability, so that
\sigthree\ and \sigstar\ agree in the mean.  This agrees
with {Boroson} (2003), who in effect uses \mbh\ to infer \sigstar\ in a sample
of low-redshift radio-quiet QSOs and Seyfert 1 galaxies in the Sloan
Digital Sky Survey (SDSS).
\footnote{The SDSS Web site is http://www.sdss.org/. }
The limited range of \mhost\ in our sample implies a limited range of 
$\sigma_*$, and the expected increase in \sigthree with \sigstar is obscured.
{Boroson} (2003) likewise found that the scatter in 
\sigthree\ artificially flattened the slope of the \mbhsigthree\
relationship in his QSO sample. 

One possible systematic error in the values of \mhost\ would be the presence
of a young stellar population associated with the AGN 
episode and any triggering galactic merger.   {Nelson} {et~al.} (2004)
find their sample of Seyfert 1 galaxies to be 0.4 magnitude brighter than
expected for the normal Faber-Jackson relation. A similar offset was found by
{Nelson} \& {Whittle} (1996).  In contrast, {McLure} \& {Dunlop} (2001) find no
offset.  {Nolan} {et~al.} (2001) examined the ages of QSO host galaxies with the aid of
off-nuclear spectra.  Their sample includes 6 RQQ and 8 RLQ
in common  with this paper. They find a predominantly old stellar
population with any young population limited to less than 1\% of the
old, or less than $\sim$0.3\% if the fits include nuclear continuum.
The stellar  population synthesis models of Charlot \& Bruzual (1991)
give an absolute magnitude per solar mass of $M_R = 3.3, 7.1$ for an
age $10^8, 10^{10}$~yr, respectively.  For a young population of age
$\sim 10^8$~yr with 0.3\% of the mass of the old population, 
the brightening is $\sim 0.1$~magnitude. Even if the brightening were
as large as the 0.4 magnitude found by Nelson et al., the difference
in expected \sigthree\ would be only  0.04 dex, not enough to seriously
alter Figure 1. 

\subsection{Does \sigthree track \sigstar?}
\label{subsec:sigsig}

Given the large scatter observed in \sigthree the question arises
if \sigthree\ does indeed track \sigstar.  Results for Seyfert and
radio galaxies do show an overall increase of \sigthree\ with
\sigstar\ or host luminosity of the expected magnitude
({Nelson} \& {Whittle} 1995, 1996).  We examine the increase 
in \sigthree\ with \sigstar\ over a wide range of AGN luminosity, deriving
\sigstar\ by various means in different ranges of luminosity,
extending the results of {Nelson} \& {Whittle} by using an
inferred \sigstar. \fref{fig:sigsig} shows \sigthree\ versus \sigstar\
deduced from a variety of sources.  For high luminosity QSOs, we
calculate \sigstar\ from the \mbhsigstar\ relation  ({Tremaine} {et~al.} 2002):
\be
\mbh =
(10^{8.13}~\msun)\left(\frac{\sigstar}{200~\mathrm{km~s}^{-1}}\right)^{4.02}.
\ee
High luminosity objects are included from 
{Shields} {et~al.} (2003), using their three highest redshift data sets, 
and near IR spectra from  {Shemmer} {et~al.} (2004) and {Netzer} {et~al.} (2004).  We
include those  
objects from the sample of  Netzer et al. and Shemmer et al. for which
their ``fit'' and ``direct'' measurements of the \oiii\ FWHM  agree
within $20 \%$ of each other in 
order to exclude those objects which do not have well measured
\oiii\ widths. 
For our data in Table 1, we convert host galaxy magnitude
to \sigstar\ via the Faber-Jackson relation. The effect of recent star
formation on the expected \sigstar\ is discussed above; this applies
only to the \sigstar\ values derived  
from host galaxy luminosity (this paper). For Seyfert and radio
galaxies, direct measurement of \sigstar\ are available in the
literature.  Nelson \& Whittle (1995) give 
values of \sigstar\ and \sigthree\ for a number of Seyfert galaxies,
noted as `Nelson 95' in  
Figures 2 and 3.  Figure 2 shows only broad line
objects, but in Figure 3 we include mean values for the Nelson \&
Whittle Seyfert 2 galaxies, noting as narrow line radio galaxies
(NLRG) Seyfert 2s with radio luminosity
L$_{1415} > 22.5$, following Nelson \&
Whittle.  All broad line radio galaxies in Nelson and Whittle
are included in Figure 2 from the later sources.  Figure 5 of
{Nelson} \& {Whittle} (1996) shows that the 
Seyfert 1 and 2 galaxies occupy a similar location in a \sigthree -- \sigstar 
diagram. In addition to objects from {Nelson} \& {Whittle},
we have taken values for \sigstar\ from {Nelson} {et~al.} (2004),  
{Onken} {et~al.} (2004), {Treu}, {Malkan}, \& {Blandford} (2004) and {Bettoni} {et~al.} (2001). FWHM of \oiii\
were obtained from {Nelson} (2000), and {Heckman} {et~al.} (1981). For objects
in {Nelson} \& {Whittle} (1995) also contained in {Nelson} {et~al.} (2004) or 
{Onken} {et~al.} (2004), we used the latter sources. \oiii\ was
measured directly from SDSS spectra for the Seyfert galaxies in
{Treu} {et~al.} \fref{fig:sigsigmeans} shows the means of
each  data set, both for RL and RQ objects. It is clear that,
though considerable scatter is present, in the mean,
\oiii\ tracks \sigstar\   over a wide range of object
luminosity. A possible shift in \sigstar\ of less than 0.1 dex due to,
e.g. brightening from young stellar populations, is again not large
enough to seriously alter \fref{fig:sigsigmeans}.

\subsection{Is [O~III] narrow in Radio-Loud Quasars?}
\label{subsec:narrower}

There is some evidence that FWHM \oiii\ in radio-loud objects tends to
be narrower than in radio-quiet objects at a given host
luminosity. \fref{fig:plot1} shows 
that the radio-loud objects in our sample have, on average, narrower
FWHM \oiii\ for a given host luminosity. We find average values of  
log \sigthree, \mhost $(2.34 \pm 0.03, -23.23 \pm
0.07)$ for radio-loud objects, and $(2.37 \pm 0.04, -22.47 \pm 0.15)$
in radio-quiet objects. (Errors are standard deviations of the mean).  
In our sample, the RL hosts are 0.76 magnitudes brighter, on average,
than the RQ hosts. Therefore, they should have log \sigthree about 0.08
larger.  In fact, we find them to have log \sigthree smaller by 0.04,
giving a total discrepancy of $0.12 \pm 0.05$ in log \sigthree with
respect to the Faber-Jackson relation.

{Shields} {et~al.} (2003) found radio-loud objects to have, on average,
narrower FWHM \oiii\ than radio-quiet objects for a given black hole
mass. {Bian} \& {Zhao} considered the \mbhsig\ relationship, taking
\mbh\ from the velocity of the broad line region (BLR) and the BLR size -
luminosity relation of  {Kaspi} {et~al.} (2000), and using published FWHM \oiii\
measurements from {Marziani} {et~al.} and
{Shields} {et~al.} They find \mbh\ to be  
larger than expected from \wthree in radio-loud objects.
Table 3 of {Bian} \& {Zhao} gives the magnitude of this offset as
$\Delta$ log \mbh = 0.51, -0.36 for RL, RQ objects in {Marziani} {et~al.}, 
respectively, and 0.59, 0.17 for {Shields} {et~al.}, where
$\Delta$ log \mbh = log (\mbh/$M_{\mathrm{[O~III]}}$) and
$M_{\mathrm{[O~III]}}$ is the black hole mass as implied by the
\mbhsig\ relation given in \sref{subsec:sigsig}. This
gives an  average $\Delta$~log \mbh(RL-RQ) of 0.65. Using $\mbh \propto
\sigma^4$ ({Tremaine} {et~al.} 2002), we may restate this as $\Delta$ log
\sigthree\ (RL-RQ) = 0.16, close to our value of
0.12.  QSOs in the SDSS Data Release 1 have [O~III]
systematically narrower for RLQ than for RQQ  at fixed \mbh\ by $\sim$
0.08 dex (Salviander et al. 2004). 
Thus there is some evidence for a RL--RQ offset at moderate QSO
luminosities. However, in Figures~\ref{fig:sigsig} and
~\ref{fig:sigsigmeans}, high luminosity 
QSOs of Shields et al. and the low luminosity Seyfert and radio
galaxies show no significant RL-RQ offset. Further complicating
the picture,  {Nelson} \& {Whittle} have found that AGN with
powerful, linear radio sources sometimes have \wthree\ larger than
expected for the value of \sigstar.

{Bian} \& {Zhao} (2004) consider whether the RL-RQ offset seen in
the \mbh\ - \sigthree\ relation might be due to the measurements of \mbh\
or \sigthree. Geometrical effects in RLQ might affect the observed
width of the broad \hbeta\ line 
or the optical continuum luminosity, either of which would affect the
derived \mbh. Alternatively, the RL-RQ offset could be
due to narrower \oiii\ lines in RL objects.
\fref{fig:plot2} shows an \mbh-\sigthree\ plot for our objects, where \mbh\ is
derived as described above. The RL-RQ offset is similar to that in the 
\mhost-log \sigthree\ plot.  \fref{fig:plot3} shows an \mbh-\mhost\ plot
along with the relationship predicted by combining the Faber-Jackson
relationship in Figure 4 of {Bernardi} {et~al.} (2003b) with the
\mbh-\sigstar\ relationship of Tremaine et al. (2002). 
The F-J relation was adjusted from $r^*$ to $R_C$ using
{Fukugita} {et~al.} (1995). \fref{fig:plot3} shows no significant RL-RQ offset
relative to the expected slope. These results suggest
that narrower \sigthree\ for RL objects is responsible for the RL-RQ offset 
in the \mbh-\sigthree relationship, and not any systematic effect
involving \mbh.  The cause of this offset is unknown.

\subsection{Do PG Quasars Have Star Formation Commensurate with 
Black Hole Growth?}
\label{subsec:growth}

The proportionality of black hole mass and bulge mass,
$\mbh \approx 0.0013 M_{\mathrm bulge}$ ({Kormendy} \& {Gebhardt} 2001) raises
the question of whether black hole growth and bulge growth occur
simultaneously.  Massive amounts of star formation are  observed in
some luminous AGN ({Sanders} \& {Mirabel} 1996).  Does
the black hole growth resulting from the accretion that powers the PG QSOs of
our sample (Table 1) correspond to a young stellar population consistent with
$\Delta \mbh \approx 0.0013~\Delta M_{\mathrm bulge}$?  Our RQQ have
an average bolometric luminosity $L \approx 10^{45.7}~\mathrm{erg~s^{-1}}$
for a bolometric luminosity estimated as $L \approx 
9~\nu~\L_\nu(5100~\mathrm \AA)$ ({Kaspi} {et~al.} 2000).  This corresponds to an
accretion rate $\dot M \approx 1~M_\odot\,\mathrm{yr ^{-1}}$ for an
efficiency $L \approx 0.1\dot M c^2$.  The corresponding star formation rate 
is $\sim 700~M_\odot \mathrm{yr^{-1}}$ to maintain the \mbhbulge\ relationship.
Rates of this magnitude are observed in some 
ULIRGs but not in these PG QSOs.  Thus the instantaneous star formation
rate is insufficient to build bulge in proportion to black hole.  However,
one may ask if a period of star formation over the recent past
may have augmented the bulge in the correct proportion to the
the growth of the black hole expected during the current QSO outburst.
As discussed above, {Nolan} {et~al.} (2001) find that any young stellar
component is not more than about 0.3\% of the old bulge stellar
component.  QSOs in our sample have $L/L_{\mathrm Ed} \approx 0.2$,
for \mbh\ and $L$ as described above. For
$L = 0.1\dot M c^2$, this gives a black hole growth time of $10^{8.4}$~yr.
Dunlop et al (2003) find an activation fraction for low redshift
RQQ of $\sim10^{-3}$, implying a minimum QSO lifetime of
$\sim10^7~\mathrm{yr}$. This would entail black hole growth by at
least $\sim4\%$, much larger than the 
fractional addition of bulge stars allowed by {Nolan} {et~al.}
Star formation in PQ quasars seems to be far less than required to
maintain detailed balance between bulge and black hole mass; however,
the main growth of the black hole at higher redshifts may involve more
nearly simultaneous star formation and bulge growth.

\acknowledgments
EWB was supported at the University of Texas at Austin by a NASA GSRP
fellowship and is supported at l'Observatoire de Paris by a
Chateaubriand fellowship. EWB thanks R. Matzner at 
the Center for Relativity for his support and encouragement. 
GAS and SS were supported under Texas Advanced Research Program grant 
003658-0177-2001 and NSF grant AST-0098594. RJM acknowledges the
support of the Royal Society. We thank K. Gebhardt,
E. Hooper, B. Wills, and T. Boroson for helpful comment.
\bibliography{}

\end{document}